# Why there is something rather than nothing
The finite, infinite and eternal

Peter Lynds[1]

**Abstract**

Many believe that the deep question of "why is there something rather than nothing?" is unanswerable. The universe *just is* and no further explanation for its existence is possible. In this paper I explain why this question must have an answer, and why that answer must establish that physical existence is inescapable and necessary. Based on the conclusion that if the universe is eternal rather than having a beginning some finite time in the past, the universe *has to* exist rather than not because its possible non-existence is never an option, such an explanation is put forward. As a logical extension of only an eternal universe being capable of providing an answer to the question of why there is something rather than nothing, the argument necessitates that the universe must be eternal. The consequences of this conclusion for cosmology are then briefly discussed.

## 1. Introduction

Why is there something rather than nothing? As Stephen Hawking put it, "What is it that breathes fire into the equations and makes a universe for them to describe?…Why does the universe go to all the bother of existing?"[2] Since this question was first posed as such by Gottfried Leibniz in 1697 [1], it has often been termed the ultimate question. Martin Heidegger called it "the first of all questions,"[3] the British astrophysicist A.C.B Lovell observed that it raised problems that could "tear the individual's mind asunder,"[4] while Charles Darwin deemed it "beyond the scope of man's intellect."[5] Despite its standing as a problem, one Adolf Grünbaum has termed the primordial existential question (PEQ) [2], it has attracted comparatively little attention from modern philosophers or physicists. This is no doubt at least partly due to it being viewed by many as being unanswerable, whether due to being seen as unfathomable, or as a pseudo-problem that does not stack up as a valid question in the first place. Due to its apparent unanswerability by conventional means, theists have claimed that the only other possible explanation is the creation of all physical existence by God [3]. However, not only do I think it a genuine problem, but also that it must have a logical, naturalistic answer.

After providing a summary of other approaches to tackling the question (placing a particular emphasis on ideas about quantum creation from nothing), in this paper I argue that such an answer is to be found through the conclusion that, contrast to a universe with a beginning a finite time in the past, only an eternal universe can possibly provide an answer to the PEQ, because at no stage during its eternal lifetime is its non-existence ever an option. I then argue that, because of this and an inescapable need for a necessary explanation for existence (contrast to a contingent one or no explanation), the universe must also be eternal. The implications of this conclusion for cosmology and theology are then discussed.

---

1  21 Seaview Road, Paremata, Wellington 6004, New Zealand. Email: peterlynds@xtra.co.nz

2  Hawking, S. *A Brief History of Time*. Bantam, New York. p. 192. (1988).
3  Heidegger, M. *An Introduction to Metaphysics*. Yale University Press, p. 1. (2000).
4  Lovell, A.C.B. *The Individual and the Universe*. Harper, New York, p. 125. (1961).
5  Darwin, C. Letter to N. D. Doedes. *Darwin Correspondence Project Database*. Letter no. 8837. (1873).

## 2. Denying non-existence

The implicit assumption underlying the PEQ is that the existence of the physical universe is *contingent.* Things may have been different and the universe not have existed. If one were able to reject this assumption, the conclusion that the universe must exist and could not have possibly done otherwise naturally follows; its existence would be *necessary*. Furthermore, because in order to answer the PEQ, we need an explanation as to why the universe exists rather than not, and by its very nature, a contingent explanation already admits the possibility of the universe not existing, any such explanation must fall short for it cannot tell us why non-existence was indeed not the case. As such, only an appeal to the necessary and non-contingent nature of existence can potentially provide a satisfactory explanation.[6]

From Parmenides onwards, denying that non-existence is possible has had a long and venerable history in philosophy, and in attempting to address the PEQ, and this is the approach that authors such as Armstrong [4], Lowe [5], and Rundle [6] have taken. However, and while it is logically contradictory to deny the existence of something because that "something" must exist for its existence to be denied, this only applies to the abstract "idea" of that something—something which does not physically exist. One can certainly assert without contradiction that a living unicorn does not physically exist, but that the idea of it exists in a non-physical, timeless, platonic sense. Furthermore, the possibility of the universe not actually physically existing—of there literally not being anything physical at all—does at least seem to be a coherent idea (even if, in modal logic terms, the idea of a "possible world" in which nothing exists is contradictory due to there not being anything existent to make such a possible world existent). After all, as Baldwin [7] and Rodriguez-Pereyra [8] have argued, anything that one may argue remains physically existent in such a context can simply then have its existence subtracted.

Contrast to Leibniz, who argued that non-existence was more simple and natural, and so required the existence of a necessary God to create a hugely complex and unlikely contingent physical existence [9], Van Inwagen has attempted to explain why there is something rather than nothing by showing not that non-existence is impossible, but existence far more probable [10]. Because there are so many ways in which there could have been something, but only one in which there is nothing, something is more probable. Indeed, given there is an infinite number of ways there could have been something, but only one in which there is nothing, nothing, while not strictly impossible, is maximally improbable.

A third approach, such as that of Max Tegmark [11] and Dean Rickles [12], is to argue that mathematical structures exist necessarily, and given that theoretical physics represents the physical universe as a mathematical structure, the physical universe exists necessarily as a consequence of mathematical structures necessarily existing. As Rickles argues:

> Either existence is contingent or it is necessary. If it is contingent then there is no complete coherent account of existence. If it is necessary then we need a necessary structure to ground this fact. Mathematical structures are of this kind. If reality is mathematical then it must exist. Reality is mathematical (as evidenced by the effectiveness of mathematics in the sciences).

---

6 This should not be confused with the question of why certain contingent physical things, such as rocks or people, exist rather than not, as these are separate questions and relate to the existence of particular things and not existence itself. Why people exist rather than not is explainable causally by physics, biochemistry and evolution, while the existence of rocks is explainable causally by geology. As we will see shortly, existence itself cannot be explained causally, however, because such an explanation will always leave the existence (rather than not) of the causal agent unexplained.

Therefore, there is existence.

However, and while it is arguably a different case for a non-physical, platonic reality,[7] the idea that *physical* reality is mathematical is extremely difficult, if not impossible to justify. While many would accept that mathematical structures exist necessarily in a platonic sense, and very few would deny the effectiveness of mathematical structures in describing and mapping the world, it can be argued that this is so simply because the world naturally has quantities, and mathematics is, by its nature, quantitative. Geometry corresponds so well to the world because the world has extent, and by default, is geometric and has dimensions. Indeed, the only way in which the world wouldn't be geometric is if it didn't exist. Given this, it should then be no surprise that mathematics and mathematical structures correspond so well to Nature. Trying to assign physical reality to mathematical structures is also deeply problematic when one considers the contingency of specific physical structures that are said to correspond to necessarily existing mathematical structures. They are not the same entities.

Finally, a fourth approach—or rather a group of related approaches—are scientific based ones, and all share the common claim that the universe spontaneously sprang into being from nothing (or at least, from something that is very close to nothing). This story started in 1973 when Edward P. Tyron published a paper asking if the universe could have arisen from a gigantic quantum fluctuation in a "pre-existing quantum vacuum" [13]. According to quantum theory, a vacuum is not actually an "empty space," but is filled with a sea of "virtual" particles that randomly fluctuate into existence for very short periods before disappearing. In a way, they are able to do this by borrowing energy from the vacuum (in the form of "vacuum energy," which is the lowest possible energy value a system can have and hypothesised to exist everywhere in the universe), and as long as these particles pay the energy back very quickly with their disappearance, they do not violate conservation laws. Despite their name, such particles are very real (if short lived), they can interact with each other, and with the Casimir effect [14], their existence has been experimentally confirmed. Taking this a step further, Tyron speculated that before the big bang, there could have been such a fluctuation, and with the help of gravity (which, representing negative energy, would offset the positive energy of the fluctuation so that energy was still conserved), an enormous random fluctuation could have resulted in the birth of the universe.

However, as Tyron's model was dependant on the existence of a background space, physicists quickly realized that this would then leave the existence of the background space unexplained. In an effort to get around this and do away with anything pre-existing, in 1982 Alexander Vilenkin put forward a model based on another well established aspect of quantum mechanics called quantum tunnelling. Although I will avoid going into details, in his model Vilenkin proposed that the universe was created by the tail of its wave function tunnelling from truly nothing [15]. Another out of nothing idea, which shares traits of both Tyron's and Vilenkin's proposals, is occasionally said to be realizable simply through thinking about what the uncertainty principle would imply about nothingness [16]. As the energy of nothing would be exactly zero, and the uncertainty principle forbids such exacting knowledge of a system, "something" in the form quantum fluctuations would inevitably result. In both scenarios, the question of why there is something rather than nothing would presumably be answered by the laws of quantum mechanics.

---

7 See, for example, the Quine-Putnam indispensability argument. H. Putnam, "Philosophy of Logic," reprinted in *Mathematics, Matter and Method: Philosophical Papers, Volume 1*, 2nd edition, Cambridge: Cambridge University Press, pp. 323-357, (1979), and, W. V. Quine, "On What There Is", reprinted in *From a Logical Point of View*, 2nd edition, Cambridge, MA: Harvard University Press, pp. 1-19, (1980).

However, there would appear to be fatal flaws with both of these ideas. If we ask the question, would the laws of physics exist in the absence of anything physical?, the answer here is clearly no, because they couldn't physically exist, and could only possibly do so in a platonic sense. Yet, with both scenarios, the laws of quantum mechanics are somehow expected to exist in the absence of anything physical to cause quantum fluctuations or tunnelling. That is, the laws are said to be causal to both, and yet, in such a scenario, these laws couldn't exist. Indeed, by appealing to these models and assuming that non-physically existing, eternal, and timeless laws of physics could somehow bridge to the physical to cause fluctuations or tunnelling, a person is actually ascribing to a belief almost perfectly analogous to the idea that the universe was created by a non-physically existing, eternal, and timeless God, who is also somehow able to bridge this non-physical–physical gap to create a universe from nothing. Moreover, to say that the energy value of nothing is zero is something of a misnomer, because nothing couldn't have an energy value. It has no properties at all, and this is what makes it nothing.

Of course, talking about the existence of the laws of physics being dependant on the physically existent is not entirely accurate, as, in a strict sense, the laws of physics don't actually exist; they are simply models that we use to describe and make predictions about the world. Yet, the physical phenomena that they describe, such as magnetism, would appear to be very real, and in the same way that magnetism can be said to exist, so too can the laws of physics. Interestingly, some physicists have argued that the laws of physics may be not be fundamental at all, but instead emergent from some more fundamental process [17], or the result of a single overriding principle [18]. Although I find this a fascinating possibility, this would not help get spontaneous quantum creation from nothing off the hook, because such processes or principles could also not have any bearing in the absence of anything physical.

Recently, the physicists Victor Stenger [18], Frank Wilczek [19], and Lawrence Krauss [20], have come up with some new arguments as to why something should spontaneously appear from nothing, and posited that physics can indeed answer the question of why something exists rather than not. These physicists all share the claim that the existence of "something" would be inescapable, because true nothingness would be perfectly symmetric (completely invariant under transformation), and because symmetric systems tend to decay to less symmetric and more complex ones, nothing would be "unstable." That is, through spontaneous symmetry breaking, "nothing" would tend to spontaneously decay into "something." But there are some serious problems with this idea too. It is difficult to see how nothing could be treated as symmetric, because nothing has no properties. For obvious reasons, one also couldn't run "nothing" through a transformation, while if one tried to, one would inevitably invoke a space. Some physicists would probably contend that one should be able to treat "nothing" just as anything else in physics—with a model—and that because one can hypothetically remove all matter, fields, and energy from a space, such an empty space or true vacuum should suffice as a model for nothing. But as this would still represent a space (a free space), one certainly wouldn't be dealing with "nothing." Indeed, as a physical space is entirely dependant on the existence of matter and energy, physics—the study of the physically existent—could no more model "nothing" than a frog.

These three physicists (ones I should note I only hold respect for!) also share the claim that a universe with total energy equal to zero (through the positive energy of matter and energy being conserved by gravity) supports the idea that the universe may have spontaneously appeared from nothing. However, as the total energy of a universe that didn't spontaneously appear from nothing would also be conserved and be zero, this doesn't really add up.

In relation to physics being able to answer the question of why something exists rather than not, one can now see why a causal explanation for the question will never work, because it will always leave the existence of the causal agent unexplained. That is, unless an explanation for existence stops with a non-causal, necessary explanation, the question cannot be answered. In the case of spontaneous creation from nothing, the explanation for existence ends with the existence of symmetry principles or the laws of quantum mechanics, which are both said to be causal to "something." But if we ask why these things would exist rather than not, considering that they could only have bearing in relation to the physically existent, we are left grasping at thin air (or nothing). Moreover, if we perhaps said that our universe *was* randomly created from nothing (perhaps from one of countless quantum flucutations out of nothing, with no background space or eternally pre-existing vacuum energy), with such a scenario, it is also possible that our universe may *not* have been created. As our universe would then represent a contingent universe, quantum creation from nothing couldn't answer the question of why *our* universe exists rather than not (nor could it in respect to any other similarly contingent universe in a possible "multiverse.")

If I had been asked one year ago what I believed the answer to the PEQ to be, I would have responded that, with no possible appeal to a causal explanation, the universe just is, and that I didn't think any further answer was possible. However, I have since realized that the only reason I would adhere to such a "brute force" explanation is simply because I couldn't imagine what a proper answer, even just in principle, might look like. The following simple argument with a logical twist is the reason for this change of mind.

## 3. The Argument

*i.* Only an eternally existing universe (defined as a state of affairs where at least one physical thing *always* remains existent) can answer the question of why the universe exists rather than not, because the alternative—a universe with a beginning a finite time in the past—presents the possibility that the universe need not have existed. By implication, before the universe's so-called beginning, there was nothing, and given that such a model posits that nothingness is possible, it is also possible that this state of affairs may have continued, with such a universe not coming into being. Given that the existence of such a universe is contingent, it cannot possibly offer a satisfactory explanation as to the question of why the universe exists rather than not, because it cannot answer the secondary part of the question, "why not nothing?".

On the other hand, however, at no time during its lifetime can an eternal universe not exist because, by definition, its existence is eternal. At no stage during its eternal lifetime is its non-existence ever an option. As such, an eternal universe would exist rather than not because it is logically contradictory for an eternal universe to ever not exist. Its existence would be *necessary (*or in the context of temporal logic, true *always*), with it being impossible for this proposition to not be true under any condition. This is the key point of this paper, and although simple, historically it puzzlingly seems to have gone unnoticed.

*ii*. Because only an eternal universe (contrast to a finite *ex nihilo* one) can answer the question of why the universe exists rather than not, and this question must have an answer, the universe must also be eternal.

*iii.* Because the universe must be eternal, the universe exists rather than not because at no stage during its eternal lifetime is its non-existence ever an option.

In light of the above argument,[8] it becomes apparent that the faulty assumption underlying the PEQ has been in our not making a differentiation between an eternal universe and a universe with a beginning a finite time in the past, and secondly, in presupposing that the possibility of the universe not existing is a valid one. While it is for a universe with a beginning a finite time in the past, this is not the case with an eternal universe.

It might be objected that it not be necessary that the PEQ be answerable (as necessitated by *ii*.). Could it be that there is no explanation? If this were so, the existence of the universe would be contingent (if it were necessary, there would be an explanation automatically given in order to ground its necessity). But contingency in this sense demands an explanation; if something is one way but could have been different, there must be an explanation for its being that way and not different. As outlined earlier, however, in the case of existence itself, no satisfactory contingent explanation is possible due to such an explanation already admitting the possibility of the universe not existing, and so not being capable of telling us why non-existence was indeed not the case; with such a universe, there could be nothing rather than something. Consequently, and even though, of itself, there doesn't seem to be any other issue with the existence of a contingent universe, one is forced into the need for a necessary explanation for existence (and a necessarily existing universe) even if one wishes to claim that no satisfactory explanation (via a contingent universe) is required.

In relation to it being contradictory for an eternal universe to not exist at some point during its eternal lifetime, it might also be contended that the same could be said for anything. For example, "It is contradictory for an eternal rock to not exist at some point during its eternal lifetime." Logically, this a valid statement. Irrespective of science telling us that rocks cannot be eternal, this would seem to spell real trouble for the argument that only an eternal universe (contrast to a finite *ex nihilo* universe) can provide an answer to the PEQ, because an eternal rock would represent another logically valid option. Indeed, we could replace a rock with anything else we might care to invoke, and each woud be equally logically valid. However, whatever we replace an eternal universe with in the argument, it would still actually be one and the same with an eternal universe. That is, an eternal rock would actually *be* an eternal universe, and we see that an eternal universe would still be the only other option or counter example to a finite *ex nihilo* universe. Of course, one could question if, like a rock, it is physically possible for the universe to be eternal. But while it may be meaningful to, the possible validity of the argument is unaffected because it is not dependant on whether an eternal universe is physically possible or not. Moreover, the argument is not concerned with what exists in the universe (this is where science comes in), but rather with showing *logically* that something, whatever it may be or how short lived, must always remain physically existent.

## 4. Discussion

The conclusion that the universe must be eternal naturally has some implications. Firstly, it would appear to preclude any cosmological model that posits that physical existence had a beginning a finite time in the past, whether that model have a finite or infinite future. As long as existence is not thought to actually "begin" at the big bang, however, such a conclusion is entirely compatible with big bang theory. It is the "before the big bang" that is the issue here, not the big bang itself. Not withstanding what the argument means for finite *ex nihilo* cosmological models, the idea of physical existence having a beginning a finite time in the past has another problem. Irrespective of the concepts of time and space having no meaning before a so-called "beginning" due to there literally being nothing there (including clocks or rulers, real or theoretical), it is still valid to question what caused that beginning. But whatever physical event one may claim caused such a universe to exist, that event must have a cause too, as must that event, and so on, and so on, adfinitum. Because of this need for an infinite regress of prior causes, such a model inevitably results in impossible contradiction. Due to it being caused by the laws of physcs, note that this problem does not apply to the theory that our universe may have arisen out of nothng from a gigantic quantum fluctuation. However, as outlined earlier, rather that requiring an infinite regress of prior physical causes, this would require that the laws of physics exist in order to cause quantum fluctuations, and apart from perhaps just in a platonic sense, they couldn't in such a situation.

---

8 The argument can also be stated as follows: (1) The idea of an eternally existing universe not existing at some point during its eternal lifetime is contradictory; logically, such a universe cannot fail to exist and so would exist necessarily. (2) The existence of a universe which begins from nothing a finite time in the past is contingent. (3) The only way to satisfactorily answer the PEQ is with a necessary explanation. (4) Only an eternal universe (contrast to a universe which begins from nothing a finite time in the past) can answer the PEQ. (5) The PEQ must have an answer. (6) The universe must be eternal.

However, the alternative option—that of the universe having an infinite past—is equally contradictory, for if the past were infinite with no beginning point from which the universe could begin to evolve from, it would be impossible for the universe to evolve to where we find ourselves today. Indeed, with an infinite past, the universe could not evolve forward (in a manner of speaking) at all. After all, at any stage in the history of the universe, time would be infinite in the direction of the past. To better illustrate this point, it may help to imagine a ruler with one end being infinite in length and the other finite, and then ask how it would be possible to arrive at the finite end when coming from the infinite or opposite direction? Hopefully one will be able to see that this would be impossible. Or to put it yet another way; as an infinite cannot actually physically pertain, and a universe with a series of events stretching infinitely into the past would represent an infinite actually physically pertaining, the past cannot be infinite. It might be contended that with a universe with an infinite past, any event would still be a finite time in the past. While this is true, any event in the past would also have an infinite number of events preceding it! It is probably worth noting that this problem is also not dependant on time existing, as the same argument could be made by not referring to tense or the past, and instead simply referring to a numerical series of events. Thus, as with a universe with a finite past, any cosmological model that posits that physical existence has an infinite past (for example, an eternally pre-existing quantum vacuum, eternal inflation [21], the Steinhardt–Turok cyclic model [22]) also results in contradiction.

But this would all appear to present an impossible situation. How is there a universe when the seemingly two only options for its lifetime, finite or infinite, both result in contradiction? Moreover, what good is the main argument of this paper if the idea of an eternal universe is so clealy fundamentally flawed? However, there is actually a third, lesser-known option. In a universe in which time is cyclic, the universe is without beginning or end, exists eternally, and yet, in relation to time is also finite.[9] I should note, however, that if one perhaps disagrees with the fore-mentioned cyclic model (considering that it posits a closed universe, most at this point![10]), or disagrees that there is a problem with the past stretching back infinitely, it does not matter in regard to the central (independent) argument of this paper. Indeed, if taken alone, the argument would compliment and support any model that posits an infinite past.

---

[9] For more details, please see, P. Lynds, On a finite universe with no beginning or end, *arXiv Preprint Server*, arxiv:physics/0612053, (2006).

[10] I should note that there are closed models in the literature that are consistent with current data [23], while I also don't see any reason why a flat universe should necessarily preclude a universe in which time was cyclic.

Before finishing, I should mention that the argument also has some significance for theology. With there being no place for a Creator in an eternally existing universe, and such a universe, of itself, already providing an answer to the PEQ, there can also be zero room for theistic explanations for why the universe exists rather than not. As the seeming unanswerability of the PEQ by any conventional means has represented something of a last bastion for physical theology, this may not go down very well. Indeed, some non-theists may also be disappointed by such an answer to the problem, for the unanswerability of the PEQ seems to add a certain mystery to existence, evoking a sense of wonder and awe in people. Because of this, some may prefer that it actually be unanswerable. I do not share this view. That we find a universe in which meaningful physical questions, even the deepest ones, all seem to have meaningful, naturalistic answers—a cosmos in which there is no need to appeal to supernatural explanations for why the world is the way it is—is, to me, a much better reason to feel a deep sense of wonder and awe when contemplating the nature of physical existence.

Thank you to Dean Rickles, Paul Halpern and Stuart Presnell for very valuable comments regarding the contents of this paper. A further thank you to Fran Healy and John Moffat, the existence of whom has encouraged mine.

[1] Leibniz, G. (1697). *De Rerum Originatione* [On the Ultimate Origin of Things].
[2] Grünbaum, A. (2004). The Poverty of Theistic Cosmology. *The British Journal for the Philosophy of Science*, 55(4): 561-614.
[3] Swinburne, R. (2004). *The Existence of God,* 2d ed., Oxford University Press.
[4] Armstrong, D. M. (1989). *A Combinatorial Theory of Possibility.* Cambridge University Press.
[5] Lowe, E. J. (1996). Why is there anything at all? *Aristotelian Society Supplementary Volume 70*: 111–20.
[6] Rundle, B. (2006). *Why there is Something rather than Nothing?* Clarendon Press.
[7] Baldwin, T. (1996). There might be nothing. *Analysis* 56: 231–38.
[8] Rodriguez-Pereyra, G. (1997). There might be nothing: the subtraction argument Improved. *Analysis*, 57, 159–166.
[9] Leibniz, G.W. *Principles of Nature and Grace*, Philosophical Papers and Letters, Leroy Loemkev Edition, University of Chicago, 1956, Vol. II, p.1033-4.
[10] Van Inwagen, P. (1996). Why Is There Anything at All? *Proceedings of the Aristotelian Society*, 70: 95-110.
[11] Tegmark, M. (2008). The mathematical universe. *Foundations of Physics* 38: 101-50.
[12] Rickles, D. (2009). *On Explaining Existence*. 2009 FQXi essay contest.
[13] Tryon, E. (1973). Is the universe a vacuum fluctuation?, *Nature* 246: 396-397.
[14] Casimir, H. B. G. (1948). On the attraction between two perfectly conducting plates. *Proc. Kon. Nederland. Akad. Wetensch*. B51, 793.
[15] Vilenkin, A. (1983). Birth of inflationary universes. *Phys. Rev. D* 27, 2848-2855.
[16] Gefter, A. Existence: Why is there a universe? *New Scientist*. Issue 2822, 16 July, 2011.
[17] Wheeler, J, A. (1990). "Information, physics, quantum: The search for links" in W. Zurek (ed.), *Complexity, Entropy, and the Physics of Information*. Addison-Wesley.
[18] Stenger, V. (2006). *The Comprehensible Cosmos: Where Do the Laws of Physics Come From?* Prometheus Books.
[19] Wilczek, F. (2008). *The Lightness of Being: Mass, Ether, and the Unification of Forces*. Basic Books.

[20] Krauss, L. (2012). *A Universe from Nothing: Why There Is Something Rather than Nothing*. Free Press.
[21] Vilenkin, A. (1983). The birth of inflationary universes. *Phys. Rev.* D27 (12): 2848.
[22] Steinhardt, P. J., Turok, N. (2007). *Endless Universe*. New York. Doubleday.
[23] Kamionkowski, M., Toumbas, N. (1996). A Low-Density Closed Universe. *Phys. Rev. Lett.* 77, 587–590.